\documentclass[11pt,twoside]{article}

\usepackage{asp2004}
\usepackage{epsf}
\usepackage{psfig}
\usepackage{lscape}

\begin{document}
\title{Future continuum surveys} 
 \author{Andrew W. Blain}
\affil{Astronomy, 105-24 Caltech, Pasadena CA91125, USA}

\begin{abstract}
A significant population of distant sub-millimeter-selected galaxies (SMGs) 
with powerful dust continuum emission that matches the luminosity of the 
brightest QSOs and exceeds that of most extreme local galaxies detected by {\it IRAS}, has 
been known for almost a decade. The full range of powerful ground- and 
space-based facilities have been used to investigate them, and a good deal of 
information about their properties has been gathered. This meeting addresses 
some of the key questions for better understanding their properties. While 
continuum detection is relatively efficient, a spectrum is 
always required both to determine a distance/luminosity, and to probe astrophysics: 
excitation conditions, the total mass, the 
mass distribution and degree of dynamical relaxation. Once a redshift is 
known, then the associated stellar mass can be 
found, and more specialized spectrographs can be used to search for 
specific line diagnostics. The first generation of submm surveys, have yielded a 
combined sample of several hundred SMGs. Here we discuss the size and follow-up of 
future SMG samples that will be compiled in much larger numbers by JCMT-SCUBA-2, 
{\it Herschel}, {\it Planck}, LMT, ALMA, and a future large-aperture (25-m-class) 
submm/far-IR wide-field ground-based telescope CCAT, planned to operate at a Chilean 
site even better than ALMA's. The issues concerning placing SMGs in the context of their environments and other populations of high-redshift galaxies are discussed.
\end{abstract}

\vspace{-0.5cm}
\section{Introduction}

\subsection{Skewering the high-redshift universe for monster galaxies} 

The existence of high-redshift far-IR-luminous galaxies was demonstrated 
using the 15-m JCMT in the 1990's (e.g Isaak et al.\ 1994). 
The detection of redshifted thermal continuum emission from interstellar dust, 
with a spectral index $\alpha \simeq 3.5$ -- where the spectral energy 
distribution (SED) follows $\nu^\alpha$ -- is possible in several mm/submm-wave 
atmospheric windows, and multiband observations rule out a substantial 
contribution from non-thermal synchrotron emission. 

The first unbiased surveys through the 450/850-$\mu$m atmospheric windows were made 
using the breakthrough JCMT-SCUBA 37-pixel array camera (Holland et al. 1999), 
starting in 1997 (Smail, Ivison \& Blain 1997). The most luminous dusty galaxies were 
found in very narrow pencil beams (with survey dimensions of order $1 \times 3000$\,Mpc). 
In principle, objects beyond re-ionization could be detected (Blain et al.\ 2002), 
but the majority appear to be at redshifts $1 < z < 3$ (Chapman et al.\ 2005). The 
geometry of these surveys remains a series of radial porcupine spikes, less than a 
total are of a square degree in extent (Laurent et al.\ 2006; Mortier et al. 2005). 

The identification of similar populations of dusty far-IR-emitting objects on the short 
wavelength side of the dust emission peak that lies at of order 100\,$\mu$m in the 
restframe was made by the ISOCAM camera (Altieri et al. 1998) at 
15\,$\mu$m, yielding a cosmologically distant ($z \simeq 1$) view of the class of galaxies 
detected by {\it IRAS}. {\it Spitzer Space Telescope} is now finding a hugely greater 
number of far-IR dominated galaxies at moderate and high redshifts. By comparison with 
SMG surveys the {\it Spitzer} surveys are much wider, 
covering about 60\,deg$^2$ to a distance in excess of $z \simeq 1$. 

Deployments of other instruments at mm/submm wavelengths, including the 1.2-mm 
MAMBO at the IRAM 30-m telescope at Pico Veleta, Bolocam at CSO, AzTEC at JCMT, 
and the forthcoming 450/850-$\mu$m SHARC-2 at JCMT and 870-$\mu$m LABOCA at the 
APEX 12-m telescope will soon swell the number of detected 
SMGs well into the thousands, while {\it Spitzer} surveys have cataloged 
more than $10^5$ far-IR objects in relatively wide, deep surveys. 
SED measurements using the 350-$\mu$m CSO SHARC-2 camera are now being 
generated for existing samples (Kovacs et al.\ 2006; Laurent et al.\ 2006). 

In Fig.\,1 the ranges of luminosities and dust temperatures associated with detectable 
SMGs and far-IR counterparts are shown for two representative redshifts $z=1$ and 3. 
The K-correction from redshifting the steep thermal Rayleigh--Jeans spectrum ensures
that the submm bands longer than about 200\,$\mu$m are very efficient to survey for 
the most luminous galaxies with only a weak dependence on redshift. However, 
the peak frequency of the dust emission spectrum, which is set predominantly by a dust temperature value ($T_{\rm d}$), also has a substantial effect on the 
detectability of SMGs. For a given luminosity and redshift, cooler SEDs can be 
more easily detected (Blain, Barnard \& Chapman 2004). 

The sensitivity levels plotted in Fig.\,1 represent the confusion limits of existing and 
future telescopes, which are the ultimate limits to the depth of their surveys, 
imposed by spatial power fluctuations from faint unresolved sources in the telescope beam. 
Confusion limits the reach of existing surveys to only the most luminous galaxies. Confusion noise leads to a beam-to-beam fluctuation level that 
is approximately the same as the flux of the source that occurs at a surface 
density of one per beam; however, the fluctuation distribution is skewed to high
flux values, and so the reliable identification of detections requires an 
intrinsic flux density of order 10 times greater (see Blain et al.\ 2002). 

\subsection{Identifying and studying the detected galaxies} 

The key problems for identifying and studying SMGs has been their significant redshifts and 
modest positional accuracy. The increase in atmospheric noise with frequency ensures that higher-resolution, shorter-wavelength submm images are rarely available and have been of only 
modest help in making identifications. Even with the most significant detections with CSO's 350-$\mu$m SHARC-2 camera, the target galaxies are still only detected in a 9-arcsec beam, 
matched to a 70-kpc scale -- much larger than the intrinsic extent of the galaxies. 
 
For the future, neither space-borne nor ground-based facilites will be able to provide 
very accurate positions and image internal structures. The CARMA, IRAM and SMA interferometers can contribute, but only in very long exposures (e.g. Tacconi et al.\ 2006). While the {\it Spitzer}, {\it Herschel} and {\it Planck} 
space telescopes can detect many of these objects, it is difficult to identify the 
most distant and luminous based on submm/far-IR detections alone
Accurate location, and size measurement will be just 
within the capability of a 10--20-m-class far-IR space telescope (e.g {\it SAFIR}), 
and an ideal goal for a space-based interferometer (e.g. {\it SPECS}). 
ALMA will be a key facility for resolving and revealing the astrophysics of 
these objects, providing rapid, spatial resolved images of SMGs in several bands, including 
a resolved spectroscopy capability. Nevertheless, ALMA will image galaxies only 
one by one, limiting the number of faint sources that can be imaged to 
the range 10-100\,hr$^{-1}$. 

\setcounter{figure}{0}
\begin{figure}
\plotfiddle{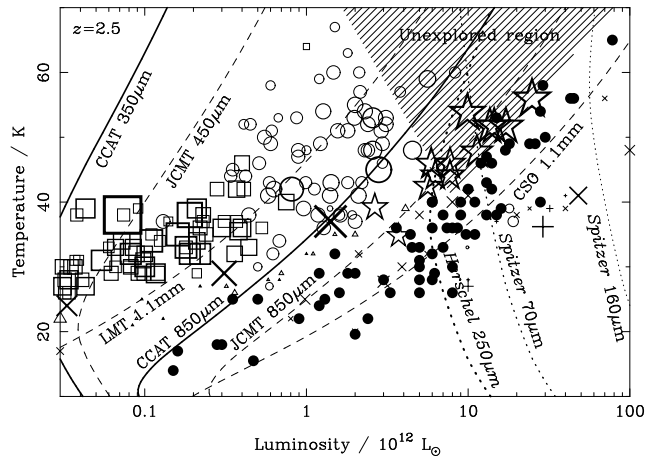}{5.8cm}{-90}{35}{35}{-130}{200}
\caption{The range of SEDs and luminosities of dusty galaxies identified at 
a range of redshifts (see Blain et al.\ 2003 for data), along with some additional 
{\it Spitzer} space telescope results (five-pointed stars). Even space-based 
facilities have difficulty in probing the shaded region. The overplotted curves 
define the confusion limits to detecting 
galaxies at $z=2.5$ with a variety of facilities. Note that submm imaging is 
crucial to detect the range of known SEDs. Only CCAT can reach luminosities 
comparable to the Milky Way ($\simeq 6 \times 10^{10}$\,L$_\odot$) 
at high redshifts.
}
\end{figure}

\subsection{Identification enabled spectroscopy} 

Throughout the initial studies of SMGs the bottleneck was obtaining redshifts, 
to both enable luminosities to be derived accurately, and via more detailed 
spectroscopy, at both near-IR and mm/submm wavelengths provide better diagnostics of 
their physical conditions. 

A modest of number of redshifts were identified serendipitously, starting with 
Ivison et al.\ (1998), who identified a galaxy in the background of the rich cluster 
A370. This galaxy has turned out to be a reasonably typical, if optically relatively bright, 
example of the population. With a redshift, information about physical 
conditions was provided by CO spectroscopy, first with the OVRO Millimeter Array  
and then the IRAM interferometer at Plateau de Bure (PDB). Recently, 
painstaking high-resolution PdB observations have allowed the first insight into 
the internal motions of mmolecular gas fueling these galaxies (see 
Tacconi et al.\ 2006 and references therein). 

A more systematic approach has been to locate the likely counterpart using 
wide-field, ultradeep radio images, enabling more certain spectroscopy. 
This relies on the detection of non-thermal radio counterparts to the SMGs, 
exploiting the accurate radio astrometry to aim multiobject optical spectrographs. 
The deepest radio images are required, in which case about 60-70\% of the SMGs are 
identified (see Chapman et al.\ 2005; this volume). 

The radio-led approach has enabled a majority of the SMGs to have precise redshifts 
assigned, allowing the subsequent investigation of the targets using millimeter, X-ray 
and near-IR spectrographs (e.g. Alexander et al. 2005; Swinbank et al.\ 2005; Hainline, this volume). These observations do not yet provide resolved 
velocity--position information to reveal the detailed astrophysical conditions within 
the SMGs. However, the SMGs can be catagorized securely as 
ultraluminous, massive, strongly clustered galaxies that  
containing relatively weak AGN (Alexander et al.\ 2005). Without redshifts, these 
investigations would either be impossible, or much more expensive in observing time. 

These follow-up observations typically require 2-3\,h of optical spectroscopy using a 
10-m-class telescope, with a significant multiplex gain at source densities of 
order 1\,arcmin$^{-2}$. They also rely on 20-h duration radio images, to enabling 
20-h duration CO observations. Hence, the amount of time required to 
carry out this process for a sample of many thousands of galaxies would be 
a great challenge. Colors from combined optical, near-IR and {\it Spitzer} imaging 
cannot be used to easily to pre-select the most likely 
counterpart to a submm detection, owing to the large surface density of candidates. 

\begin{figure}
\plotfiddle{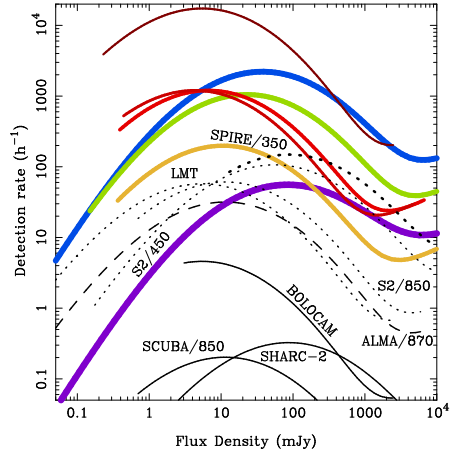}{5.8cm}{-90}{38}{38}{-135}{205}
\caption{Comparative detection rate of galaxies using labelled 
facilities/wavelengths and CCAT (at 200, 350, 450, 620, 740, 850\,$\mu$m and 
1.1\,mm in order of decreasing line width). At faint fluxes the curve for each device is 
truncated at an aggressive definition of the confusion limit.}
\end{figure} 

\section{Future continuum surveys} 

The biggest surveys so far (Mortier et al.\ 2005; Laurent et al.\ 2005), provide a clear 
picture of the nature of the majority of the luminous SMG population when combined with 
suitable spectroscopy (Chapman et al.\ 2005). However, the time expended in compiling a sample of order 100 optical spectra has required of order 20 nights of time using the Keck telescopes. 
Hence, the current follow-up process {\emph{cannot}} be scaled up directly 
to cope with the fruits of the next generation of surveys. A process to sift, prioritize and classify the detected galaxies based on their submm properties will be important. 

This detection rates estimated in Fig.\,2 provide an estimate of the difficulty of obtaining a large number of redshifts for galaxies found using JCMT-SCUBA-2, APEX-LABOCA, 
LMT-AzTEC or ALMA. The situation is even more challenging for surveys using the 
proposed Cornell--Caltech Atacama Telescope (CCAT). CCAT is a 25-m-class telescope with a 
20-arcmin field of view, operating at wavelengths as short as 200\,$\mu$m from a mountaintop close to the ALMA site. Current first-light instrumentation for CCAT includes a short 
wavelength camera, consisting of 32,000 pixels and a long wavelength camera, making 
multicolor 2\,mm-670\,$\mu$m images using an array of $\simeq 16$,000 microwave-addressed 
MKIDS detectors, which produce a prodigious detection rate. If next-generation wide-field instrumentation can be accommodated, LMT also the capability to make very 
large mm-wave surveys. CCAT's multi-color submm surveys will sample the SED where the 
thermal spectrum starts to peak, and so include significantly more information about the 
source population than detections of pure Rayleigh-Jeans thermal emission.\footnote{This could be even more important if additional emission components can be found, such as ices in pre-reionization objects (Dudley, Imanishi \& Maloney 2006)}

While the clustering and colors of the SMGs that these surveys will uncover remain interesting, 
the key questions that need to be answered about their nature all rely in some way on 
spectroscopic follow up. 

\section{Key questions for understanding the SMG population} 

In the light of existing surveys, there are two main outstanding questions. 
First, what is the nature of the significant minority of the existing samples that 
have not yielded redshifts via the radio--optical route? Secondly, what is the 
relationship between SMGs and other samples of high-redshift galaxies? These 
questions require additional capabilities in the mm/submm: the identification of 
a larger fraction of the SMGs in the surveys, and the detection of 
less luminous examples. Natural capabilities to help with these
questions are the subjects of this meeting, `z-machines' and ALMA respectively. 

The precision and flux density range of source counts, color distributions, and 
possibly angular correlation functions will be better established as the SMG sample sizes 
increase. Current counts extend from about 0.5--20\,mJy, 
with an accuracy of order 10\%, while little is known about mm--submm colors 
(Laurent et al.\ 2006; Kovacs et al.\ 2006). Where redshift and 
color information coincide, the SEDs of the SMGs are consistent with a range of 
temperatures spread by about 30\%, centered on 40\,K (Kovacs et al.\ 
2006) -- remarkably close to the temperature inferred by assuming that 
both the scant sample of galaxies from the first-generation surveys from {\it ISO} and SCUBA in 
1998 were drawn from the same strongly-evolving population (Blain et al.\ 1999).   

Without three-dimensional redshift information, determining clustering properties in 
projection with such deep surveys is hard -- the 3000\,Mpc depth of an SMG survey corresponds 
to a 110-deg swath of sky to match its width and depth at $z=3$, while to enclose even a 
single realization of the minimum representative comoving volume that is matched to the 
largest scale observed in the galaxy distribution today (of order 10$^6$\,Mpc$^3$) 
requires a field about 0.4\,deg$^2$ in size for $1<z<3$.

Hence, covering a large area of the sky in an SMG survey is essential, both to map the 
evolving structure that could be highlighted by a population of SMGs biased to 
the densest regions, and to identify the most extreme examples of the population, 
including the 10-20\% of 100-200\,mJy galaxies at 850\,$\mu$m (at a density of order
0.1\,deg$^{-2}$) that are likely to be gravitationally lensed by foreground 
galaxies (Blain 1998). 

\subsection{Links to other populations: a factor of 10 greater depth}

The kind of survey discussed above with SCUBA-2 will generate a 
larger samples of SMGs that are similar to or more extreme than those known currently,
and not a significantly less luminous 
population. They thus cannot readily detect galaxies that bridge between large 
spectroscopic samples of optically-selected galaxies (Steidel et al.\ 2003) and the 
rarer, more luminous SMGs. The limit to their depth is imposed by the floor of confusion 
noise (Fig.\,1). As it is difficult to determine the degree of dust extinction present 
from even the most extensive optical, near- and mid-IR photometry; however, 
far-IR/submm data remains an essential supplement in order to identify the true 
luminosity of high-redshift galaxies, and to construct a complete and accurate luminosity function.  

While ALMA can reach unprecedented depths, it can 
only do so in modest field areas owing to its sub-arcminute instantaneous field of view. 
Ultradeep pencil beam surveys will be ALMA's unique territory, and it is 
ideal for imaging any known or discovered target of arbitrary brightness. 
Moreover, being capable of unique spectroscopic imaging. Nevertheless, ALMA is not best suited to surveying a representative volume of 
the high-redshift Universe for typical $L^*$ galaxies, owing to the need to expend large 
amounts of time that would be better devoted to detailed imaging (see Fig.\,2). This is 
due to the modest field of view of ALMA. 

While ALMA cannot survey rapidly, 
existing single-antenna facilities are unsuitable for beating confusion and to reach the 
necessary depths to feed typical galaxies to ALMA. Focal plane MMIC arrays on interferometers 
like CARMA, combined with giant FPGA-based programmable correlators, could provide a bridge to 
survey the relevant population at mm wavelengths. A 25-m-class submm telescope with excellent 
optical quality at an excellent site -- like CCAT -- could also detect and locate typical 
galaxies directly in great numbers for follow-up spectroscopic imaging using ALMA. Working at 
submm wavelengths, CCAT would also provide SED information for the detected 
galaxies. 

The reason that CCAT 
is so much more capable than a 10-15-m class submm telescope, or a 50-m mm-wave 
telescope is made clear by the confusion noise levels represented in Fig.\,1.
Owing to the steep distribution of source flux densities in the submm waveband, and to the turn 
over in the counts at a flux density of order 1\,mJy in order to match the measured 
finite background radiation intensity (Fixsen et al.\ 1998; Kashlinsky et al.\ 2005), the 
level of confusion noise drops dramatically as aperture increases beyond 15-m and observing 
wavelength is reduced shortwards of 1\,mm. When combined with the greatest atmospheric 
transparency, this is the key to the scientific power of CCAT, securely based on the direct results of existing observations.   

\subsection{To define a true 3-D distribution - complete spectroscopy, over 
a representative volume}

To survey a volume with a cubic geometry, the existing `hypodermic needle' 
beams of submm surveys must be extended to 100-deg scale fields. While a less extreme 
pencil-beam geometry would still provide a good view of large-scale structure, a 
distance estimate is still required for a substantial fraction of the $2 \times 10^5$ 
1-mJy galaxies expected over a 100-deg$^2$ field. The capability of deep optical redshift 
surveys are increasing dramatically over the next decade. An instrument like the now-defunct WFMOS for 
Gemini\footnote{http://www.noao.edu/meetings/subaru/} will be capable of producing 4000 redshifts h$^{-1}$ at an efficiency similar to existing spectrographs that can obtain 
30-40 in the same time. The patrol field of WFMOS was 2\,deg$^{-2}$, ideally matched to the density of mJy SMGs. Hence, optical spectroscopy of the 
detected galaxies would be a very practical proposition. Although optical spectroscopy 
is likely to miss about 50\% of SMG targets, and accurate positions are still required for 
fiber positioning, substantial spectroscopical coverage for both SMGs and their 
companions and common structures will be available for next-generation surveys. However, it 
will be a stretch to keep pace with CCAT's potential 2$\pi$ sky coverage to 1\,mJy at 
350 and 850\,$\mu$m in of order 10\,yr. It will be feasible to assign 10--100-h exposures to 
target the faintest galaxies, given that 4000 galaxies could be targeted simultaneously with an instrument like 
WFMOS. The difficulty of identifying the most extreme, distant and potentially exciting targets 
found at submm wavelengths is difficult to assess, but SMGs are known to be Ly-$\alpha$ 
emitters (Chapman et al.\ 2005), a trend that should increase at high redshifts as 
metallicity declines and morphology is likely even more irregular (Chapman et al.\ 2003), a spectrograph covering to 0.3--1.0-$\mu$m can detect lines out to $z=7.2$, and it 
is unlikely that a more than a few percent of SMGs lie beyond reach. 

A very deep imaging survey from a LSST, including PanSTARRS could provide a route to 
identifying many SMGs are known to be extended, patchy and with a wide range of colors, both 
internally and across the population. Combined with the power of the 
eVLA\footnote{http://www.aoc.nrao.edu/evla} to obtain $\mu$Jy sensitivity maps over a 
1-deg$^2$ field in about a day, and relying on sub-arcsec-accuracy centroids 
from detections at greater than 10$\sigma$ levels in 
350-$\mu$m CCAT maps, there is an excellent opportunity to follow-up future SMG surveys. 

\subsection{Pre-reionization SMGs -- first metals and earliest structures}

The most exciting dusty galaxies are any found during or prior to 
reionization, subject to the assumption that metallicity and dust content is 
sufficiently great. These assumptions are reasonable following the detection of some 
of the most distant SDSS QSOs at mm wavelengths. 
These galaxies are no more difficult to detect in the 
continuum at submm wavelengths 
than those at $z \simeq 1$, but are much harder to detect at any other wavelengths. 
ALMA can certainly detect and image any such galaxies, but it will be difficult to 
identify, pre-select and determine their redshifts from imagingi data alone. 

\section{The role of photometric redshifts} 

\subsection{Optical/near-IR} The industry of obtaining redshift estimates from the 
colors of stellar populations can be used to provide estimates for SMG counterparts. 
While subject to uncertainty about the correct identification, and possible chance superpositions, the results of a deep LSST survey are likely to be very useful for providing information about large-scale structure traced by SMGs. 

\subsection{Far-IR/submm/radio} 

The huge haul of galaxies that will be found out to redshifts far beyond unity in future 
suirveys can be identified with a variety of archived data from the space missions 
{\it Spitzer}, {\it Herschel}, along with the {\it ASTRO-F}, {\it Planck Surveyor} 
{\it WISE} all-sky surveys, and an eVLA (and ultimately SKA) large-area FIRST-style 
survey program. With benign assumptions about the range of SEDs for SMGs (Blain et al.\ 2003) 
it is possible to obtain photometric estimates of redshifts for SMGs from (Carilli \& 
Yun 1998; Aretxaga et al.\ 
2005; this volume), and of course owing to the thermal nature of the SED,
very accurate measurements of $T/(1+z)$ can be obtained. 
Surprisingly, the very different emission mechanisms responsible for the far-IR--radio 
correlation does not break this temperature--redshift degeneracy, leading to a 
similar measurement of $T/(1+z)$ from radio--submm colors (Blain 1999; Yun \& Carilli 2002). 

While the information on this combined $T$-$z$ quantity is valuable, no increase in the 
number of data points or the photometric accuracy can overcome the degeneracy, unless 
there is prior information 
available for a tight relationship between the luminosity and temperature of the target 
sources. The best indications are that the temperatures of SMGs are scattered by of 
order 30\% (Laurent et al.\ 2006; Kovacs et al.\ 2006). 
(Carilli \& Yun 1998; 

With ALMA, or a z-machine, the spectral index of the SMG thermal emission can be 
measured directly in the band, to provide a very similar redshift indicator, as shown in 
Fig.\,3. This provides similar information to a color measurement, but requires
only a single tuning. As with color determination, the accuracy of the fitted 
SED information is greater if the peak of the SED is probed at submm wavelengths. 

\subsection{Radio spectral details} 

Multi-color radio observations may provide redshift probes, as the reddest radio 
sources are likely to be found where the more intense high-redshift CMB provides 
a more efficient loss mechanism for relativistic electrons than the internal 
magnetic fields of galaxies, especially prior to reionization. While this spectral 
index will not provide an accurate redshift, it might be a sufficient guide to 
highlight the galaxies with the faintest, reddest and 
steepest radio spectral indices that could be the best targets for 
detailed follow-up study as potential pre-reionization examples of the population. 

\begin{figure}
\plotfiddle{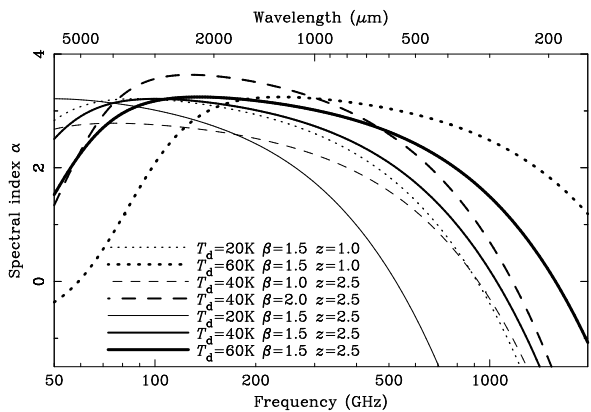}{5.3cm}{-90}{35}{35}{-130}{185}
\caption{The spectral index $\alpha$ ($f_\nu \propto \nu^{\alpha}$) as a function 
of frequency for different SEDs, defined by temperature $T_{\rm d}$, emissivity 
index $\beta$ and redshift. ALMA and redshift machines can both determine the 
continuum spectral index in a single tuning. The index can be 
used as a redshift indicator, subject to the quantity $T/(1+z)$ being 
determined accurately, as for submm/mm colors. The change in spectral index with 
redshift--temperature is more dramatic at shorter submm wavelengths. 
}
\end{figure}

\section{The role of ALMA} 

ALMA will be capable of imaging any galaxy detected at long wavelengths in a matter 
of minutes to hours, and is almost immune to confusion noise. Hence, obtaining a 
spectrum from ALMA in a 8-GHz band will allow a CO line to be detected in about 
25\% of tunings on a chosen target at $z \sim 2.5$ (CO ladder separation of 
$115/(1+2.5) = 33$)\,GHz. ALMA can thus likely find redshifts for even the most difficult 
cases after four adjacent tunings, and can do it even more easily for higher redshifts:
at reionization ($z=6.7$), adjacent CO lines are separated by just 15\,GHz.

\subsubsection{Line to continuum ratio} 

By determining the equivalent width (EW) of the CO line, either ALMA or a z-machine 
spectrograph can assign a likely transition to a single line (Blain 2000): owing to the 
redshifting of the peak of the underlying continuum emission, the CO EW should decrease 
systematically as $J$ increases. After detecting one line, this estimate of $J$ obtained 
from the EW should thus enable a direct search using ALMA, choosing the frequency tuned 
to detect an expected confirming CO line with a known $J$ value in an available band. 

\section{Summary} 

\begin{enumerate}
\item The capability of continuum detectors at mm/submm wavelengths is becoming 
dramatically greater. SCUBA-2 nears reality, while several new concepts 
are being demonstrated for multiplexing  
detectors and filling large focal planes with them. A large-aperture, wide-field 
submm-wave telescope like CCAT would be the ideal platform for them: able to overcome 
confusion noise and probe a large fraction of the SED for accurate determination of 
luminosity. 
\item The capability of ALMA and next-generation optical spectrographs are sufficiently 
powerful that spectroscopy of galaxies detected in the submm should be viable, 
with a WFMOS-like spectrograph finding redshifts for over 1000 candidates per hour, and ALMA studying the 
astrophysics of many tens per hour. Hence, multiwaveband follow up is definitely 
going to remain important. 
\item Redshift machines will provide valuable astrophysical information for exciting 
difficult cases in the period before ALMA is in service. Line to continuum ratios, 
and luminosity--temperature constraints can be exploited to help interpret the results. 
\end{enumerate}

\acknowledgments The research underlying this contribution rely on observations 
made in collaboration, especially with Scott Chapman, Ian Smail and Rob Ivison. The 
CCAT study is led at Cornell and Caltech by Terry Herter and Jonas Zmuidzinas 
respectively, with project management by Tom Sebring and Simon Radford. 
I thank the Alfred P. Sloan Foundation and the Research Corporation for support.

\vfill\pagebreak

\end{document}